\documentclass[letterpaper, 10 pt, conference]{ieeeconf}  % Comment this line out if you need a4paper

\usepackage{amssymb}
\usepackage{mathtools}
\usepackage{amsmath}
\usepackage{verbatim}
\usepackage{soul}
\usepackage{ifthen}
\usepackage{xkeyval}
\usepackage{xcolor}
\usepackage{tikz}
\usepackage{calc}
\usepackage{graphicx}
\usepackage{fancyheadings}
\usepackage{booktabs}
\usepackage{multirow}
\usepackage{cite}
\usepackage[normalem]{ulem}
\IEEEoverridecommandlockouts                              % This command is only needed if 
\newtheorem{theorem}{Theorem}

\newtheorem{remark}{Remark}
\newtheorem{proposition}{Proposition}
\title{Minimum-Energy Distributed Filtering% <-this % stops a space
\thanks{This work was supported by the Australian Research Council under Discovery Projects funding scheme (project D0120102152).}}

\author{Mohammad Zamani \and Valery Ugrinovskii% <-this % stops a space
\thanks{M. Zamani and V. Ugrinovskii are with School of Engineering and IT, 
        UNSW Canberra, Canberra, Australia, Email: 
        {\tt\small \{m.zamani,v.ougrinovski\}@adfa.edu.au}}}

\begin{document}

\maketitle
\thispagestyle{empty}
\pagestyle{empty}

%%%%%%%%%%%%%%%%%%%%%%%%%%%%%%%%%%%%%%%%%%%%%%%%%%%%%%%%%%%%%%%%%%%%%%%%%%%%%%%%
\begin{abstract}
%%%%%%%%%%%%%%%%%%%%%%%%%%%%%%%%%%%%%%%%%%%%%%%%%%%%%%%%%%%%%%%%%%%%%%%%%%%%%%%%
The paper addresses the problem of distributed filtering with guaranteed convergence properties using minimum-energy filtering and $H_\infty$ filtering methodologies. % Distributed filtering concerns the estimation of the state of a plant using a network of locally communicating sensors that share information to each produce a local estimate. 
A linear state space plant model is considered observed by a network
of communicating sensors, in which individual sensor
measurements may lead to an unobservable filtering problem. However, each filter locally
shares estimates,  that are subject to disturbances, with its respective
neighboring filters to produce an estimate of the plant state. The
minimum-energy strategy of the proposed local filter leads to a locally
optimal time-varying filter gain facilitating the transient and the
asymptotic convergence of the estimation error, with guaranteed  
$H_\infty$ performance. The filters are implementable using only the local
measurements and information from the neighboring filters subject to
disturbances. A key idea of the proposed algorithm 
is to \emph{locally} approximate the neighboring estimates, that are
not directly accessible, considering them as disturbance contaminated
versions of the plant state. % The proposed method is advantageous since it
% results in a completely decoupled design algorithm. 
The proposed algorithm imposes minimal communication load on the network
and is scalable to larger sensor networks.   

% This approximation is inspired by the fact that a similar minimum-energy filter is implemented at each node where the local estimate ideally converges to the true state, asymptotically. The approximation error is also included the minimum-energy filtering problem to establish the asymptotic convergence property of the approximated neibouring estimates.
\end{abstract}
%%%%%%%%%%%%%%%%%%%%%%%%%%%%%%%%%%%%%%%%%%%%%%%%%%%%%%%%%%%%%%%%%%%%%%%%%%%%%%%%
\section{INTRODUCTION}
%%%%%%%%%%%%%%%%%%%%%%%%%%%%%%%%%%%%%%%%%%%%%%%%%%%%%%%%%%%%%%%%%%%%%%%%%%%%%%%%

There is considerable interest in the literature in multi-agent systems
that are capable of performing control and filtering related tasks in a
cooperative manner. Applications range from military aerial fleets,
monitoring and maintenance agents in industrial applications to biological
applications. In this paper, in particular, we are concerned with
distributed filtering using a network of filters that estimate the state of
a plant using disturbance contaminated local measurements. An interesting
case is when these filters individually have difficulty providing an
accurate and complete estimate of the plant, a problem that is resolved by
using information from the neighbouring filters. 

Kalman filtering % ~\cite{Kalman}
is the focus of many of the existing methods
proposed for distributed filtering. An early result on this subject by
Durrant-Whyte \emph{et. al}~\cite{durrant90,Rao93} provides an exact
decentralized formulation for the multi-sensor version of the Kalman
filter. This formulation avoids the requirement of a central processing or
communication unit with which each sensor has to communicate its
information for calculating the state estimate of the plant. The
decentralized scheme is robust to sensor failure and network changes,
reduces the communication load and allows for faster information
processing. The decentralized Kalman filter proposed
in~\cite{durrant90,Rao93} however, 
requires an all-to-all communication of state error information and
variance error information between the sensors; this impairs its scalability
to bigger networks. Olfati-Saber~\cite{Olfati2005} 
proposed a distributed Kalman filter algorithm that involves two additional
consensus filters which are needed for fusion of the sensor information and
covariance information required for the decentralized Kalman
filtering scheme used. The distributed naming refers to the fact that
all-to-all information sharing is avoided and instead each agent only
shares information with its local neighbors. The distributed consensus
algorithms proposed in~\cite{Olfati2005} assume disturbance free data
sharing between the sensors that is arguably an unrealistic assumption in
applications involving imperfect communication.  
  
More recently, observer designs based on linear matrix inequalities (LMIs)
have been employed to tackle the imperfect communication problem in multi-agent
distributed 
estimation~\cite{ugrinovskii2011distributed,Subbotin20092491}. Subbotin and 
Smith~\cite{Subbotin20092491} proposed a convex optimization problem where
minimizing the estimation error covariance of the global network selects
the local observer gains.  Ugrinovskii~\cite{ugrinovskii2011distributed}
designed the observer gains by minimizing an $H_\infty$ consensus
performance cost that guarantees the associated disturbance attenuation
criterion and guarantees the convergence of the filter dynamics of
the nodes. Both works target the asymptotic performance of their respective
estimation algorithms by using constant gains. This is in contrast to
filtering algorithms that use time-varying gains 
calculated to adjust both the transient and the asymptotic performance 
~\cite{Anderson}.

In this paper a minimum-energy distributed filtering algorithm is proposed
that addresses the imperfect communication of the nodes by locally
minimizing a quadratic energy cost comprising the initialization and
measurement errors as well as the communication errors of the local
filter. A novel contribution of this paper that leads to the distributed
nature of the local filters lies in augmenting the minimum-energy cost
functional with a cost on approximating the neighboring estimates. 
%  Approximating the neighboring estimates as copies of the plant state contaminated with minimum-energy error is justified by the fact that each agent uses a minimum-energy filtering strategy that will lead to a minimum-energy type estimation error.
Furthermore, a penalty is included in the local costs that provides a
 guaranteed $H_\infty$ disturbance attenuation property over the network of
 filters. 
 
We provide a condition on the network parameters that is sufficient to
ensure convergence of the proposed filters. The condition is quite
general, and captures the types of convergence considered
in~\cite{ugrinovskii2011distributed,Subbotin20092491} as special cases. 
Moreover, we offer a tuning algorithm based on LMIs that facilitates the
implementation of the proposed filters with guaranteed $H_\infty$
performance and convergence of the 
filters. A simulation example regarding the Chua electronic circuit~\cite{Ugrinovskii2013160} is provided that demonstrates the effectiveness
of the proposed algorithm in the case of a chaotic plant system and network of  five filters that estimate the state of the plant.   

The remainder of the paper is organised as follows. Section~\ref{ME} reviews the
concept of minimum-energy filtering. In
Section~\ref{DistFilProb} we extend the minimum-energy filtering paradigm
to allow for a distributed filtering formulation as well as to impose the
$H_{\infty}$ filtering performance criterion. The distributed
minimum-energy filter is provided in that section. Convergence of the
proposed filter is studied in 
Section~\ref{stability}, and tuning of the filter is discussed in
Section~\ref{VU.tuning}. Section~\ref{sim} illustrates the design procedure
and performance of
the proposed filter via simulation, and Section~\ref{conclusion} concludes
the paper.      
 
% The important point in this formulation is that communicating neighboring state estimates subject to error in a network amounts to less load and accuracy demand on a network than the requirements of many similar works in the literature \textbf{\color{red} add citations to OlfatiSaber works and others}. Most of the mentioned works assume that perfect neighboring states are available to each sensor. Another common assumption is that each agent is able to receive the measurements of the other agents in its neighborhood that in the case of multiple measurements increases the network load. This  problem is also avoided in this work by restricting the communications to neighboring estimates only. 
%%%%%%%%%%%%%%%%%%%%%%%%%%%%%%%%%%%%%%%%%%%%%%%%%%%%%%%%%%%%%%%%%%%%%%%%%%%%%%%%
\section{Minimum-Energy Filtering}\label{ME}
%%%%%%%%%%%%%%%%%%%%%%%%%%%%%%%%%%%%%%%%%%%%%%%%%%%%%%%%%%%%%%%%%%%%%%%%%%%%%%%%
In this section we review the concept of minimum-energy filtering that was pioneered by Mortensen~\cite{Mortensen} and was further elaborated by Hijab~\cite{Hijab}. Later, we use this method to obtain a distributed filtering algorithm for estimating the state of a linear system using a network of filters. Consider the linear system 
\begin{equation}%\label{linsys}
\label{state}
 \dot{x} = Ax +Bw, \quad x(0)=x_0,
\end{equation}
where the signals $x\in\mathbb{R}^{n}$ and $w\in\mathbb{R}^{m}$ are,
respectively, the state and the unknown modeling disturbance; the latter is
assumed to be $\mathcal{L}_2$ integrable on $[0,T]$. The matrices
$A\in\mathbb{R}^{n\times n}$ and $B\in\mathbb{R}^{n\times m}$ are known
state matrix and input disturbance coefficient matrix, respectively. Note
that $x_0$ is assumed to be unknown. 

 Further consider the following measurement model
\begin{equation}\label{meas}
 y = Cx+ Dv,
\end{equation}
where the signals $y,v\in\mathbb{R}^{p}$ are the measured data and the
unknown measurement disturbance, which is also assumed to be
$\mathcal{L}_2$ integrable on $[0,T]$. The matrices $C
\in\mathbb{R}^{p\times n}$ and $D\in\mathbb{R}^{p\times p}$ are the
measurement matrix and the measurement disturbance coefficient matrix,
receptively, that are assumed to be known from the model. Denote
$Q\triangleq BB^{\top}$ and $R\triangleq DD^{\top}$ and assume that $R>0$.  
 
Associated with these equations is the energy functional measuring the
aggregated energy associated with the unknowns $(x_0,w,v)$,  
\begin{equation}\label{energyfunctional}
 J_t =  \frac{1}{2}\Vert x_0-\xi \Vert_{\mathcal{X}}^ 2 + \frac{1}{2} \int_0^t\left( \Vert w \Vert^2  +  \Vert v\Vert^2\right)d\tau,
\end{equation}
where $\Vert a\Vert_A\triangleq a^{\top}Aa$, the matrix
$\mathcal{X}\in\mathbb{R}^{n\times n}$ is a weighting on the difference
between the unknown initial state and its nominated \emph{a priori}
estimate $\xi$.  

Denote by $y_{[0,t]}$ the data obtained according to~\eqref{meas} during
the time $[0,t]$. Given the measurement data $y_{[0,t]}$, minimizing the
cost~\eqref{energyfunctional} with respect to $w$ and $x(0)$, subject to
equations~(\ref{state}) %~\eqref{linsys} 
and~\eqref{meas}, leads to an optimal state trajectory $x^*_{[0,t]}$. This
is the `most likely trajectory'~\cite{Mortensen} or the minimum-energy
trajectory that is compatible with the data $y_{[0,t]}$.  
The end point of this trajectory constitutes the minimum-energy estimate of
the state $x(t)$, given the measurement data $y_{[0,t]}$,
\begin{equation}
\hat{x}(t)\triangleq x^*_{[0,t]}(t), \quad t\in[0,T]. 
\end{equation}

 It is desirable to obtain the estimates $\hat{x}(t)$ continuously as time
 evolves in  $[0,T]$. Note that in general, for $0\leq t_1 < t_2\leq T$, 
 \begin{equation}
  \hat{x}(t_1)=x^*_{[0,t_1]}(t_1)\not = x^*_{[0,t_2]}(t_1).
 \end{equation}
 Therefore, it is not sufficient to only minimize the
 cost~\eqref{energyfunctional} over $[0,T]$ and infinite sequence of
 minimizations are required if all the estimated values in this period are
 needed. The calculus of variations can be utilized to solve this
 problem and obtain a recursive filtering
algorithm similar to those arising in optimal
control~\cite{Athans}. The idea is to solve the following two step
optimization problem  
\begin{equation}\label{optcontprob}
\begin{split}
\inf_{x(0)}(\inf_{w} J_t)=& \inf_{x(0)} \left[\frac{1}{2}\Vert x(0)-\xi
  \Vert_{\mathcal{X}}^ 2\right. \\ 
&+ \left.\inf_{w}\frac{1}{2} \int_0^t( \Vert w \Vert^2  
 +\Vert y-Cx\Vert_{R^{-1}}^2)d\tau\right].
 \end{split}
\end{equation}
The inner minimization problem is solved as an optimal tracking problem, in
which the system~(\ref{state}) %(\ref{linsys}) 
is to track the given
output signal $y(\tau)$, $\tau\in[0, t]$, and $w$ is treated as a
control input signal. For this step both
$x(0)$ and $x(t)=x$ are considered 
to be fixed but constrained by~(\ref{state}) %~\eqref{linsys} 
for $0\leq \tau\leq t$. The associated value function of this inner
tracking problem, encoding the minimum energy required to take the system
from the initial state $x(0)$ to the state $x$ over the time period $[0,t]$, is
\begin{equation}\label{Value}
 V(x,t) = \min_w J_t, 
\end{equation}
which from~\eqref{energyfunctional} has the initial condition
\begin{equation}
 V(x,0)= \frac{1}{2}\Vert x-\xi \Vert_{\mathcal{X}}^ 2.
\end{equation}
Note that in (\ref{Value}), 
the variables $x$ and $t$ are considered as independent
variables. That is, by selecting a time $t$ and an end state $x$, the
value function~\eqref{Value} provides the optimal value arising from
minimizing the cost~\eqref{energyfunctional} over $w$.  

 The final step is to perform the outer minimization, which is equivalent
 to minimizing the value function over $x$. This leads to the optimal
 trajectory $x^*$ and the minimum-energy estimate chosen as its final
 value, $\hat{x}(t)=x^*(t)$. Provided the value function is sufficiently
 smooth, this can be achieved by solving  
\begin{equation}\label{valuef}
\nabla_x V(x,t) = 0.  
 \end{equation}

Mortensen~\cite{Mortensen} proposed solving the next equation, obtained from~\eqref{valuef}, that leads to a recursive set of equations for updating $\hat{x}(t)$.
\begin{equation}\label{Mortensen}
\dfrac{d}{dt} \{ \nabla_x V(x,t)\} = 0.
 \end{equation}
The resulting filter turns out to be in the exact form of the Kalman-Bucy
filter~\cite{Anderson}. While the resulting filter is well-known in the
literature, the minimum-energy methodology is less known. Nevertheless,
minimum-energy filtering is a systematic recursive method that can be
applied in many applications including in distributed filtering considered
in the rest of this paper.  

%%%%%%%%%%%%%%%%%%%%%%%%%%%%%%%%%%%%%%%%%%%%%%%%%%%%%%%%%%%%%%%%%%%%%%%%%%%%%%%%
\section{Minimum-Energy Distributed Filtering}\label{DistFilProb}
%%%%%%%%%%%%%%%%%%%%%%%%%%%%%%%%%%%%%%%%%%%%%%%%%%%%%%%%%%%%%%%%%%%%%%%%%%%%%%%%

Consider a network of $N$ filters with a directed graph topology
$\mathbf{G}=(\mathbf{V},\mathbf{E})$ where $\mathbf{V}$ and
$\mathbf{E}\subseteq \mathbf{V}\times \mathbf{V}$
are the set of vertices and the set of edges, respectively. An edge
directing node $j$ of the graph $\mathbf{G}$ towards node $i$ where
$i,j\in\{1,\cdots,N\}$ is denoted by $(i,j)$. In accordance with the common
convention~\cite{Olfati2005}, the graph $\mathbf{G}$ is
assumed to have no self-loops, i.e., $(i,i)\notin \mathbf{E}$. 
The neighborhood of node $i$, i.e., the set of nodes which send information
to node $i$, is denoted by $\mathcal{N}_i=\{j:(i,j)\in\mathbf{E}\}$.   

In this section we again consider the plant model  
% \begin{equation}\label{state}
% \dot{x} =  Ax + Bw,\quad x(0)=x_0,
% \end{equation}
% where the vector $x\in\mathbb{R}^n$ is the state of the plant, the matrices 
% $A\in\mathbb{R}^{n\times n}$ and $B\in\mathbb{R}^{n\times m}$ are known from the model with $Q\triangleq BB^{\top}$ and the signal 
% $w \in\mathbb{R}^m$ is an unknown modeling error.
(\ref{state}). As in Section~\ref{ME}, the initial state $x_0$ and the
$\mathcal{L}_2$-integrable modeling error $w$ are unknown. Our
goal is to construct a network of filters that can each 
estimate the plant state $x$ using the measurements of the following type,  
\begin{equation}\label{measurement}
y_i = C_ix + D_iv_i,
\end{equation}
where $i=1,2,\cdots,N$, indicates the $i$th node of the network. 
Each measurement $y_i\in\mathbb{R}^p$ is obtained using the associated
observation matrix $C_i\in\mathbb{R}^{p\times n}$ known to node $i$ from
the model. The signal $v_i\in\mathbb{R}^p$ is an unknown measurement error
assumed to be $\mathcal{L}_2$ integrable on $[0,T]$, with known coefficient
matrices $D_i\in\mathbb{R}^{p\times p}$ such that the matrix $R_i\triangleq
D_iD_i^{\top}$ is positive definite.  

In addition, the filter $i$ obtains information from $l_i$ 
agents in its neighbourhood $\mathcal{N}_i$; $l_i$ is the cardinality of
the set $\mathcal{N}_i$, $0\leq l_i\le N$. The obtained signals at node $i$ are 
\begin{equation}\label{communication}
c_{ij} = W_{ij}\hat{x}_j + F_{ij}\epsilon_{ij},\quad j\in\mathcal{N}_i,
\end{equation}
where $\hat{x}_j$ is the estimate of state $x$ at node
$j\in\mathcal{N}_i$. Similar to the measurements $y_i$, the signals 
$c_{ij}\in\mathbb{R}^m$ are obtained using the weighted connectivity
matrices $W_{ij}\in\mathbb{R}^{m\times n}$ that are assumed to be known to
node $i$ while $\hat{x}_j$ are not known. The signals
$\epsilon_{ij}\in\mathbb{R}^m$ are unknown 
communication errors of class $\mathcal{L}_2[0,T]$, with known coefficient
matrices $F_{ij}\in\mathbb{R}^{m \times m}$ such that  $S_{ij}\triangleq
F_{ij}F_{ij}^{\top}$ are positive definite. 

\begin{remark}
An interesting situation arises when some or all the pairs $(A,C_i)$ are
not detectable and hence the state $x$ cannot be estimated at the
corresponding filter nodes from 
$y_i$ only. Forming a large-scale distributed observer network by enabling the
filters to obtain information about the state estimates of their neighboring filters
allows these nodes to overcome the lack of detectability and provide a 
reliable state estimate $\hat{x}_i\in\mathbb{R}^n$, e.g.,
see~\cite{U7b-journal} where this issue is discussed in detail. It should
be noted that the distributed filtering problem under consideration is
different from a centralized filtering scheme where all the measurements
$y_i$ are sent to a central processing unit to obtain an estimate of the
plant's state. In a centralized scheme the computational complexity at the
central node, communication load on the network and scalability are the
common issues that distributed filtering aims to address.    
\end{remark}

In many existing methodologies,
cf.~\cite{ugrinovskii2011distributed,Ugrinovskii2013160,Subbotin20092491},   
treating communications between the nodes as extra measurements results
in observer design conditions that require solving large-scale coupled
matrix equations or matrix inequalities. In contrast, we invoke the following
equation that is going to be instrumental in obtaining decoupled filter
gain conditions. Since $\hat{x}_j$ are meant to represent the state vector $x$
with high fidelity, we use the following model at node $i$,
 \begin{equation}\label{xhatj}
  \hat{x}_j= x+\eta_{ij}.
 \end{equation} 
 The signals $\eta_{ij}\in\mathbb{R}^b$ denote the unknown error signals of this assignment. Note that this model is only proposed for node $i$ and node $j$ obviously computes $\hat{x}_j$ using its corresponding filter.  
 
 Next, we pose a minimum-energy filtering problem to obtain a distributed
 filter at each node $i$ such that a cost functional, depending on the
 unknown initial state of the plant $x_0$ and the unknown errors in the
 measurements and neighbouring information associated to filter $i$, is
 minimized. This cost functional is in fact the sum of aggregated energies
 in the unknowns of the model, the measurements and the neighbouring information
 associated with filter $i$. Note that the distributed estimation paradigm
 restricts us to only utilizing local measurements $y_i$ and local
 information from the neighboring nodes,
 i.e. $c_{ij},\;j\in\mathcal{N}_i$ in this optimization process. It is
 important to stress the role of the unknown signals $\eta_{ij}$ that are
 also to be optimized over in this process.

The idea behind minimizing the cost functional over $\eta_{ij}$, 
$j\in\mathcal{N}_i$, amounts to agent $i$ relying on the fact that its
neighboring agents are also minimum-energy type estimators of the plant
state and hence their estimates can be thought of as estimates of
the plant state contaminated with errors that are based on minimum-energy
filtering too.  This is a key idea among the results of this work that
facilitates obtaining a distributed solution for our problem. The concept
of minimum-energy filtering was explained in details in Section~\ref{ME}.  
  
The energy cost similar to (\ref{energyfunctional}), associated with the
system~\eqref{state}, the information available to node $i$ according
to~\eqref{measurement} and~\eqref{communication}, and the
approximation model~\eqref{xhatj} is defined as
 \begin{equation}\label{ecost}
 \begin{split}
& \frac{1}{2}\|x_0-\xi_i\|_{\mathcal{X}_i}^ 2 +\frac{1}{2}
  \int_0^t\bigg(\|w \|^2+ \|v_i\|^2  \\
&\qquad\qquad\qquad +\sum_{j\in\mathcal{N}_i}(\| \epsilon_{ij} \|^2 
+\|\eta_{ij}\|^2_{Z^{-1}_{ij}})\bigg)d\tau,
\end{split}
\end{equation}
 where the signal $\xi_i$ is an a priori candidate for the initial state $x_0$
 at node $i$ and the matrix $\mathcal{X}_i$ is an a priori known weighting
 for the initialization error $x_0-\xi_i$. The weighting matrices
$Z_{ij}=Z_{ij}'>0$, $Z_{ij}\in\mathbb{R}^{n\times n}$, quantify the
`confidence' of node $i$  
in the quality of its  neighbor
estimates $ \hat{x}_j$ and are selected locally by node
$i$.    
 
Given the measurement data $y_i\vert_{[0,t]}$ and the
communication data $c_{ij}\vert_{[0,t]}$, the signals $v_i$ and
$\epsilon_{ij}$ are identified as dependent variables and the energy cost
functional~\eqref{ecost} has arguments $x_0$, $w_i$ and
$\;\eta_{ij}$. Next, based on (\ref{ecost}) define the following cost
functional for node $i$    
\begin{eqnarray}
&&J_{i,t}(x_0,w,\{\eta_{ij}\}) = \frac{1}{2}\Vert x_0-\xi_i \Vert_{\mathcal{X}_i}^ 2 + \frac{1}{2} \int_0^t\bigg( \Vert w \Vert^2  
\nonumber \\
&&\quad+\Vert y_i - C_ix\Vert^2_{R_i^{-1}} + \sum_{j\in\mathcal{N}_i}(\Vert c_{ij} - W_{ij}x-W_{ij}\eta_{ij}\Vert^2_{S_{ij}^{-1}}\nonumber\\
&&\quad+ \Vert\eta_{ij}\Vert^2_{Z^{-1}_{ij}}-\Vert
x-\hat{x}_i\Vert^2_{M_i^{-1}})\bigg)d\tau,
\label{costr}
\end{eqnarray}
where the matrix $M_i>0\in\mathbb{R}^{n\times n}$ is a given
coefficient. The selection of this matrix will be discussed in
Section~\ref{VU.tuning}. The inclusion of the error term $\int_0^t\Vert 
x-\hat{x}_i\Vert^2_{M_i^{-1}}d\tau$ with a negative sign enforces a guaranteed
$H_\infty$ type performance at node $i$ while a minimum-energy estimate is
sought~\cite{McEneaney}.  

The principle of minimum energy requires that a set of the
unknowns $(x_0,w,\eta_{ij})$, $j\in\mathcal{N}_i$, be sought at each node $i$
that are compatible 
with the measurements $y_i$ and the neighbourhood information $c_{ij}$ in
satisfying~\eqref{measurement},~\eqref{communication} and~\eqref{xhatj} while minimizing
the cost~\eqref{costr},
\begin{equation}\label{problem}
\inf_{x}\;\inf_{w,\;\eta_{ij}} J_{i,t}(x_0,w,\eta_{ij}),
\end{equation}
where $x=x(t)$. Using the minimum energy filtering approach as  was explained in Section~\ref{ME} the following filter is obtained 
\begin{eqnarray} 
\dot{\hat{x}}_i &=& A\hat{x}_i + K_i^{-1} \bigg(C_i^{\top} R^{-1}(y_i -
C_i\hat{x}_i) \nonumber \\
&+&\sum_{j\in\mathcal{N}_i}W_{ij}^{\top}U^{-1}_{ij}(c_{ij} -
W_{ij}\hat{x}_i)\bigg),\quad\hat{x}_i(0) = \xi_i.
\label{obs}
\end{eqnarray}
The matrix $U_{ij}$ is defined as $U_{ij} \triangleq S_{ij} + W_{ij}Z_{ij}W_{ij}^{\top}$.
 The positive definite time-varying gain matrix $K_i$ is calculated from
 the following Riccati equation
\begin{eqnarray}
\dot{K}_i &=& -K_iQK_i + C_i^{\top} R_i^{-1}C_i +
\sum_{j\in\mathcal{N}_i}W_{ij}^{\top}U^{-1}_{ij} W_{ij} \nonumber \\
 &&-M_i^{-1}-A^{\top}K_i-K_iA, 
\label{Riccati} \\
\quad K_i(0) &=& \mathcal{X}_i. \nonumber
\end{eqnarray}
The matrix $Q$ is defined as $Q\triangleq BB^{\top}$ where $B$ was given in~\eqref{state}.
% The Riccati equation~\eqref{Riccati} provides the time evolution of the
%matrix $K_i$ in the observer~\eqref{obs}. The initial condition 
%\begin{equation}
% K_i(0) = \mathcal{X}_i,
%\end{equation}
%is computed using~\eqref{value function} and~\eqref{finalcondition}.

  An advantage of equation~\eqref{Riccati} is that it does not depend on
  any measurements, neighbour information or any on-line acquired data and hence
  can be fully solved off-line. It can also be solved on-line forward in
  time, so that the computed values for $K_i(t)$ can be used to dynamically
  run the observer~\eqref{obs}. Also, the filter 
  equation~\eqref{obs} has the standard form of a distributed filter that
  produces state estimates based on local measurements and information
  obtained about the neighbours. A noteworthy feature of the proposed filter
  is that its gains are dependent only on the information available at node
  $i$. The only matrix to be computed to find the gains, the matrix $K_i$
  is computed locally, without interacting with the neighbours, provided
  the matrices $M_i$ are defined a priori. This
  feature of the proposed minimum energy estimator sets it apart from other
  distributed estimation algorithms such as those proposed
  in~\cite{ugrinovskii2011distributed,Subbotin20092491}. 
\section{Convergence Analysis}\label{stability}
%=========================================================
In this section we provide sufficient conditions to guarantee that the
interconnected network of filters designed according to
equations~\eqref{obs} and~\eqref{Riccati} converges to the true state of
the 
plant as $t\to\infty$. Due to the presence of uncertainty, the convergence
is understood in the $H_{\infty}$ sense. Firstly, we show that
error dynamics exhibit properties of an internally stable system. The
notion of internal stability is equivalent to asymptotic convergence of the
estimation error of the network in the absence of disturbance
signals. Secondly, we 
will show that when disturbances are $\mathcal{L}_2$ integrable on
$[0,\infty)$, the filter ensures disturbance attenuation properties similar
to those of an $H_{\infty}$ filter.
% 
%  Furthermore, we provide conditions that guarantee an H$_{\infty}$
%  convergence of the estimates, by introducing an H$_{\infty}$ type
%  performance in the cost functional~\eqref{cost}.

Consider the node error $e_i\triangleq \hat{x}_i-x$ and its dynamics,
  \begin{eqnarray}
  &&\hspace{-5ex}\dot{e}_i = Ae_i - Bw \nonumber \\
&&\hspace{-2.5ex}- K_i^{-1}\bigg((C_i^{\top} R_i^{-1}C_i
+\sum_{j\in\mathcal{N}_i}W_{ij}^{\top}U^{-1}_{ij}
W_{ij})e_i \nonumber \\
&&\hspace{-2.5ex}-C_i^{\top}R^{-1}_iD_iv_i -
\sum_{j\in\mathcal{N}_i}W_{ij}^{\top}U^{-1}_{ij}(W_{ij} e_j+F_{ij}
\epsilon_{ij})\bigg),
\label{error_i}
  \end{eqnarray}
  where the dynamics of the matrix $K_i$ are given in~\eqref{Riccati}. 

 Let us introduce the following notation 
\begin{equation}\label{globalp}
\begin{split}
& e\triangleq[e_1^{\top}, \ldots,  e_N^{\top}]^{\top}, \\
& \tilde{A}\triangleq I_N\otimes A,\quad C\triangleq \mbox{diag}[C_1,\ldots, C_N],\quad\tilde{Q}  \triangleq I_N\otimes Q,\\
& R\triangleq \mbox{diag}[R_1,\ldots, R_N],\quad K\triangleq \mbox{diag}[K_1,\ldots, K_N],\\
& M\triangleq \mbox{diag}[M_1,\ldots, M_N], \\
& \Delta\triangleq \mbox{diag}[\sum_{j\in\mathcal{N}_1}W_{1j}^{\top}U^{-1}_{1j}W_{1j},\ldots,\sum_{j\in\mathcal{N}_N}W_{Nj}^{\top}U^{-1}_{Nj}W_{Nj}],\\
& \tilde{\Delta}\triangleq
\mbox{diag}[\sum_{j\in\mathcal{N}_1}W_{1j}^{\top}U^{-1}_{1j}W_{1j}Z_{1j}W_{1j}^{\top}U^{-1}_{1j}W_{1j},\ldots,
\\
&\quad\quad\quad\sum_{j\in\mathcal{N}_N}W_{Nj}^{\top}U^{-1}_{Nj}W_{Nj}Z_{Nj}W_{Nj}^{\top}U^{-1}_{Nj}W_{Nj}],\\
&\tilde{L}\triangleq [L_{ij}], \\
& L_{ij}=\begin{cases}
\sum_{j\in\mathcal{N}_i}W_{ij}^{\top}U_{ij}^{-1}W_{ij}Z_{ij}W_{ij}^{\top}U_{ij}^{-1}W_{ij}, \\
 & \hspace{-2cm} i=j, \\
-W_{ij}^{\top}U^{-1}_{ij} W_{ij}, & \hspace{-2cm} i\neq j,\ j\in \mathcal{N}_i,\\
0, & \hspace{-2cm} i\neq j,\ j\not\in \mathcal{N}_i,
\end{cases}
\end{split}
 \end{equation}
 where $\mbox{diag}[X_1 ,\ldots, X_N ]$ denotes the block diagonal matrix
with $X_1 ,\ldots, X_N$ as its diagonal blocks, and $\otimes$ is the
Kronecker product of matrices. Using this notation, % the evolution of
% the global error dynamics can be written as
% \begin{equation}\label{gerr}
% \dot{e} = (\tilde{A}-K^{-1}C^{\top}R^{-1}C-K^{-1}L)e -Bw+K^{-1}C^\top R^{-1}Dv,
%  \end{equation}  
%  where
 $K$ satisfies the differential Riccati equation
 \begin{eqnarray}
\dot{K} &=& 
-K\tilde{Q}K + C^{\top}R^{-1}C +
\Delta-M^{-1}\nonumber \\
&&-\tilde{A}^{\top}K-K\tilde{A}, \label{gRicc} \\
K(0)&=& \mbox{diag}[\mathcal{X}_1 ,\ldots, \mathcal{X}_N ]. \nonumber
 \end{eqnarray}

The following theorem is the main result of this paper, which characterizes
$H_\infty$ performance of the distributed minimum energy filter
(\ref{obs}).  

\begin{theorem}\label{stabt}
Given a positive semidefinite weighting matrix $P=P'\in\mathbb{R}^{nN\times
  nN}$, 
suppose a block diagonal matrix $M=M^\top>0$ is such that 
\begin{equation}\label{Minv}
M^{-1} > P-\tilde{L}-\tilde{L}^{\top} 
+\tilde{\Delta}
\end{equation}
and each Riccati equation~\eqref{Riccati} has a
positive definite bounded solution on $[0,\infty)$. Then, the
filtering algorithm~\eqref{obs},~\eqref{Riccati} guarantees the
satisfaction of the  global disturbance attenuation criterion\footnote{In
  (\ref{H_inf}), $\Vert\cdot\Vert_2$ denotes the norm in
  $\mathcal{L}_2[0,\infty)$.}     
  \begin{eqnarray}\label{H_inf}
 \int_0^\infty e^{\top} Pe d\tau &\leq & \sum_{i=1}^N \Vert x_0-\xi_i
 \Vert^2_{\mathcal{X}_i} + N\Vert w \Vert_2^2\nonumber \\
 & + &\sum_{i=1}^N \left(\Vert
   v_i\Vert_2^2+\sum_{j\in\mathcal{N}_i}\Vert\epsilon_{ij}\Vert_2^2\right).
 \end{eqnarray}
Moreover, 
%  \begin{equation}\label{assump1}
% \mbox{Ker}(C^{\top}R^{-1}C)\cap \mbox{Ker}(L)=\{0\},
%  \end{equation}
in the absence of the disturbances $w , v_i, \epsilon_{ij}$,
$j\in\mathcal{N}_i$, $i=1,\ldots,N$, the global estimation error $e$
asymptotically converges to zero as $t\to \infty$.  
\end{theorem}
%=====================================================================================
%=====================================================================================
The proof is % carried out for instance using the Lyapunov function $ \mathcal{V}_i(e_i,t)=e_i^{\top}K_i(t)e_i$. We have
omitted % the proof
due to page restrictions.  

Several special cases of Theorem~1 deserve attention. 

\paragraph*{Minimum energy filter with guaranteed $H_\infty$ convergence
  performance} 
In this special case, the performance objective of interest
is to achieve the internal stability of the
filtering error dynamics, and enforce an acceptable disturbance
attenuation performance of the filter error dynamics. Let
$P=\mbox{diag}[P_1\cdots,P_N]$, $P_i=P_i'>0$. 
Then the conditions of Theorem~\ref{stabt} specialize to guarantee the
following $H_\infty$ condition
\begin{eqnarray}\label{H_inf.1}
 \sum_{i=1}^N\int_0^\infty e_i^{\top} P_ie_i d\tau &\leq & \sum_{i=1}^N
 \Vert x_0-\xi_i \Vert^2_{\mathcal{X}_i} + N\Vert w \Vert_2^2\nonumber \\
 & + &\sum_{i=1}^N \left(\Vert
   v_i\Vert_2^2+\sum_{j\in\mathcal{N}_i}\Vert\epsilon_{ij}\Vert_2^2\right).
 \end{eqnarray}

\paragraph*{Minimum energy filter with guaranteed transient $H_\infty$ consensus
  performance} 
Consider $P=\frac{1}{2}(L+L_{\top})\otimes P_0$ where $P_0=P_0'\ge 0$, and
$L$, $L_{\top}$ are respectively the Laplace matrix of
the network graph and that of its transpose, i.e., the graph obtained from the
network graph by reversing its edges. This choice of $P$ results in the
left hand side of the~\eqref{H_inf} being equal to the weighted $H_\infty$
disagreement cost between the nodes 
$\frac{1}{2}\int_0^\infty\sum_i\sum_{j\in\mathcal{N}_i}\Vert
\hat{x}_i-\hat{x}_j\Vert_{P_0}^2d\tau $;
cf.~\cite{ugrinovskii2011distributed,Ugrinovskii2013160}. Therefore, if
$P_0$ is such that the conditions of Theorem~\ref{stabt} are satisfied,
then this choice of $P$ adjusts the filter to guarantee 
the $H_\infty$ consensus performance,
\begin{multline}\label{H_inf.2}
\frac{1}{2}\int_0^\infty\sum_{i=1}^N\sum_{j\in\mathcal{N}_i}\Vert
\hat{x}_i-\hat{x}_j\Vert_{P_0}^2 d\tau \leq \sum_{i=1}^N
 \Vert x_0-\xi_i \Vert^2_{\mathcal{X}_i}  \\
+ N\Vert w \Vert_2^2+\sum_{i=1}^N \left(\Vert
   v_i\Vert_2^2+\sum_{j\in\mathcal{N}_i}\Vert\epsilon_{ij}\Vert_2^2\right).
\end{multline}

%=============================================================
\section{Network Design and Tuning}\label{VU.tuning} 
Theorem~\ref{stabt} provides a sufficient condition guaranteeing the
convergence in the $H_\infty$ sense of the network of filters consisting of
the estimators~\eqref{obs},~\eqref{Riccati}. To apply this condition
the matrices $M_i$ need to be selected to satisfy (\ref{Minv}). 
% each node needs to have
% access to the overall disturbance attenuation performance weight $P$ and
% the matrix $\tilde{L}$ which encodes the topology of the graph. This is
% arguably a less 
% restrictive requirement than  that of the requirements of some of the other
% methodologies demanding each node to have access to the full measurements
% of the other nodes or the exact estimates and the covariances of the
% neighboring nodes or having to run coupled LMIs to compute the filter gains
% at each node.  
The following propositions are instrumental in designing these
matrices. First, we recall a connection between the differential Riccati
equation~\eqref{Riccati} and a corresponding algebraic Riccati equation
(ARE)~\cite{Basar}, 
 \begin{eqnarray}
Z_iA^{\top}+AZ_i&-&Z_i(C_i^{\top}R_i^{-1}C_i\nonumber \\
&+&[\Delta]_{ii}-M_i^{-1})Z_i+BB^{\top}=0.  
\label{ARE}
\end{eqnarray}

\begin{proposition}\label{ARE-DRE}
If the (ARE)~(\ref{ARE}) has a positive definite stabilizing solution
$Z_i^+$ then the equation~\eqref{Riccati} with an initial condition
$K_i(0)=\mathcal{X}_i\ge (Z_i^+)^{-1}$ has a bounded positive definite
solution on $[0,\infty)$ and $\lim_{t\to \infty}K_i(t)\to (Z_i^+)^{-1}$. 
\end{proposition}

% \begin{proof}
% The proposition readily follows
% from~\cite[Theorem~9.7(vi)]{Basar}\footnote{The assumption
%   of~\cite[Theorem~9.7(vi)]{Basar} that $(A,B)$ is
%   stabilizable and $(A,[C_i^{\top}R_i^{-1/2}~[\Delta]_{ii}^{1/2}])$ is
%   detectable is only used in that theorem to establish
%   the existence of a stabilizing $Z_i^+>0$. Since we assume in
%   Proposition~\ref{ARE-DRE} that such $Z_i^+$ 
%   exists, those assumptions can be dispensed with.}, which
% shows that for any $Q_f=Q_f^\top\ge 0$ such that $Q_f\le Z_i^+$, the 
% differential Riccati equation 
%  \begin{eqnarray}
% \dot Z_i&+&Z_iA^{\top}+AZ_i \nonumber\\
% &-&Z_i(C_i^{\top}R_i^{-1}C_i+[\Delta]_{ii}-M_i^{-1})Z_i+BB^{\top}=0,
% \nonumber \\
% &&Z_i(t_f)=Q_f\label{DRE} 
% \end{eqnarray} 
% has a positive definite solution $Z_i(t_f,t)$ on $[0,t_f]$ and
% $\lim_{t_f\to \infty}Z_i(t_f,t)=Z_i^+$ for all $t$. Letting
% $Q_f=\mathcal{X}_i^{-1}$ and substituting $K_i(t)=(Z_i(t_f,t_f-t))^{-1}$ into
% (\ref{DRE}) completes the proof.
% \end{proof}

Next, we use the above connection between the ARE
(\ref{ARE}) and the differential Riccati equation (\ref{Riccati}) to   
derive a tractable sufficient condition for the conditions of
Theorem~\ref{stabt} to hold.   
  
\begin{theorem}\label{tuning.lemma}
  Suppose $(A,B)$ is stabilizable, and  $M$ and $P$ are the matrices 
  defined in Theorem~\ref{stabt}. The linear matrix inequality (LMI)
  conditions  
\begin{eqnarray}
  &&\hspace{-.5cm}
\left[\begin{array}{cc}A^{\top}X_i+X_iA-C_i^{\top}R_i^{-1}C_i-[\Delta]_{ii}+M_i^{-1}&
    X_iB\\
  B^{\top}X_i&-I\end{array}\right]< 0, \nonumber \\
  &&X_i=X_i^\top >0, \quad i=1,\ldots,N, \nonumber \\
&& \tilde{L} + \tilde{L}^{\top} -\tilde{\Delta}
  +M^{-1}> P \label{LMI}   
\end{eqnarray}
guarantee the satisfaction of the conditions of Theorem~\ref{stabt} for any
sufficiently large $\mathcal{X}_i$.  
 \end{theorem}
%======================================
%  \begin{proof}
% The existence of a positive define solution $X_i$ to the first inequality
% in (\ref{LMI}), and via the Schur complement, to the algebraic Riccati
% inequality (with $Z_i=X_i^{-1}>0$)   
%  \begin{equation}\label{MRI}
%  Z_iA^{\top}+AZ_i-Z_i(C_i^{\top}R_i^{-1}C_i+[\Delta]_{ii}-M_i^{-1})Z_i+BB^{\top}<0 
% \end{equation}
% guarantees that the algebraic Riccati equation (\ref{ARE}) has a
% positive semidefinite stabilizing solution~\cite{Gahinet94LMI}
%  \begin{equation}\label{ARE}
% Z_iA^{\top}+AZ_i-Z_i(C_i^{\top}R_i^{-1}C_i+[\Delta]_{ii}-M_i^{-1})Z_i+BB^{\top}=0. 
% \end{equation}
% Furthermore, this solution is positive definite since $(A,B)$ is
% stabilizable~\cite[Theorem 4.8(ii)]{Basar}. According to
% Proposition~\ref{ARE-DRE}, this proves that each differential Riccati
% equation~\eqref{Riccati} with the initial condition $\mathcal{X}_i\ge
% (Z_i^+)^{-1}$ has a positive definite bounded solution in
% $[0,\infty)$. The condition~\eqref{Minv} is trivially satisfied
% by~\eqref{LMI}.   
%  \end{proof}
%=========================================

In the light of Theorem~\ref{tuning.lemma}, solving the LMI~\eqref{LMI}
in variables $X_i$, $M_i^{-1}$, $P$ is considered as a tuning process for the
proposed filtering 
algorithm~\eqref{obs},~\eqref{Riccati}. Tuning can 
be carried out,  e.g., using MATLAB. It can be performed
assuming the matrix $P$ is given, to obtain the matrices $M_i^{-1}$ and
$\mathcal{X}_i$ to be used in equation (\ref{Riccati}). According to Theorem~\ref{stabt}, this
tuning process will guarantee the disturbance attenuation and the internal
stability of the error dynamics for the entire networked filter consisting
of the node filters~\eqref{obs},~\eqref{Riccati}. Alternatively, the matrices
 $M_i$, $i=1\ldots,N$, can be selected a priori. The
filter gain equations~\eqref{Riccati} are completely decoupled in this
 case, and  Theorem~\ref{tuning.lemma} can be used to compute the matrix $P$ to
 characterize performance of the corresponding distributed
 filter~\eqref{obs}. 

% The conditions of Theorem~\ref{tuning.lemma} can also be instrumental for
% tuning the 
% filter~\eqref{obs},~\eqref{Riccati} in a situation where the network graph
% varies, and several network topologies need to be considered. Most existing  
% methodologies demand the filter gains to be computed every time the
% network topology changes. Often coupled LMIs need to be solved to compute
% the filter gains at each node and for each instance of the network
% graph~\cite{Ugrinovskii2013160}. The proposed filter is arguably less
% demanding. Indeed, conditions (\ref{LMI}) can be expanded to include 
% LMIs for several network topologies, so that common 
% parameters $M_i$ can be found for all expected network configurations. 
% These parameters can then be used in the
% filter~\eqref{obs},~\eqref{Riccati}. Even though the Riccati
% equations~\eqref{Riccati} need to be adjusted when the graph changes, the
% computation of $K_i$ can be continued using the new graph information and
% new initial conditions, but with the original $M_i$. Hence, the bottleneck
% of data communication associated with solving coupled design conditions can
% be reduced substantially. 

%=====================================================================
\section{Illustrative Example}\label{sim}
%=====================================================================
In this section, a simulated network of five sensor nodes is considered
that are to estimate a three-dimensional plant. The plant's state matrix is
given by 
\begin{equation}\label{A} 
 A = \left[\begin{array}{ccc}
      -3.2  &  10  &  0  \\
      1 & -1 &  1 \\
      0 & -14.87 &  0 \\
     \end{array}\right].
\end{equation}
This corresponds to a Chua electronic circuit which was considered
in~\cite{Ugrinovskii2013160}. A Chua circuit is a chaotic system switching
between three regimes of operation. Here, we focus on one of the regimes
and consider the second mode of the Chua example that was considered
in~\cite{Ugrinovskii2013160}.  
%The measurement matrices $C_i$ are chosen such that no pair $(A,C_i)$ is detectable.
The $C_i$ matrices considered are $C_1=C_4=0.001 [3.1923~ − 4.6597~ 1]$ and $C_2=C_3=C_5=[−0.8986~ 0.1312~ − 1.9703]$.
The network considered consists of five nodes
with connectivity edges
$E=\{(1,3),(2,3),(3,1),(3,2),(3,4),(4,3),(4,5),(5,4)\}$. 
Following~\cite{Ugrinovskii2013160}, all
communication matrices associated with the sensor nodes $i$ and $j$ are taken
to be $W_{ij}=I_{3\times 3}$ if the pair  $(i,j)$ belongs to the set of
edges $E$, and otherwise $W_{ij}=0_{3\times 3}$. 

We have considered the remainder of the coefficient matrices to be
$B=0.4I_{3\times 
  3}$, $D_i = 0.025I_{1\times 3}$ and $F_{ij} = 0.5I_{3\times 3}$. The
measure of confidence in the neighbouring estimates is set to be
$Z_{ij}= 0.1I_{3\times 3}$. The initial values for the Riccati differential
equations (\ref{Riccati}) 
are chosen to be $K_i(0)=10 I_{3\times 3}$ and the 
% following $H_\infty$ gain matrix is considered 
% \begin{equation}
% \begin{split}
% M^{-1} =& \mbox{diag}[2.8100,4.6869,0.6696, 0.6508,0.6501,0.6647,\\
% &1.4949,4.4042,4.0730,    0.5882,0.6575,0.7103,\\
% &0.7790,0.6816,4.2554,3.6725,0.5961,0.7098,\\
% &0.7311,0.7042,0.6500,2.8477,3.8809,0.6745,\\
% &0.4830,0.6999,0.7047,0.8668,2.5363,4.3978,\\
% &3.4785,0.6578,0.7052,0.8261,1.4944,3.1818].
% \end{split}
% \end{equation}
%  Note that this
matrix $M^{-1}$ is chosen by solving~\eqref{LMI} with the disturbance
attenuation weighting matrix
$P=\frac{1}{2}(\mathcal{L}+\mathcal{L}_\top)\otimes I$ (to ensure~\eqref{LMI}
holds strictly, in fact we solved~\eqref{LMI} with
$P=\frac{1}{2}(\mathcal{L}+\mathcal{L}_\top)\otimes I+0.01I$). According
to~Theorem~\ref{tuning.lemma}, this guarantees the satisfaction of the
conditions of Theorem~\ref{stabt}. Furthermore, with this choice of matrix
$P$, the proposed minimum energy filter guarantees that estimation error
dynamics are internally stable and exhibit the $H_\infty$ performance
expressed in condition (\ref{H_inf.2}). 
%  \begin{split}
%   &\left[\begin{array}{cc}-\tilde{A}^{\top}K-K\tilde{A}+C^{\top}R^{-1}C+\Delta-M^{-1}&K\tilde{B}\\
%   \tilde{B}^{\top}K&I\end{array}\right]\geq 0,\\
%   &K>0,\\
%   &\tilde{L} + \tilde{L}^{\top} -\tilde{\Delta} +M^{-1}\geq P = I.
%   \end{split}
% \end{equation}
% Note that the first inequality consists of the Schur compliment of the following Riccati inequality that ensures existence of solutions for the Riccati~\eqref{gRicc}.
% \begin{equation}
%  K\tilde{B}\tilde{B}^{\top}K+\tilde{A}^{\top}K+K\tilde{A}-C^{\top}R^{-1}C-\Delta+M^{-1}<0.
% \end{equation}
% Solving the conditions~\eqref{LMI} is a tuning process for the proposed
% filters that is carried out using the YALMIP toolbox~\cite{YALMIP} in
% MATLAB. This tuning process will guarantee the stability of the proposed
% filters according to Theorem~\ref{stabt}. 

\begin{figure}
\begin{center}
\includegraphics[scale=.23]{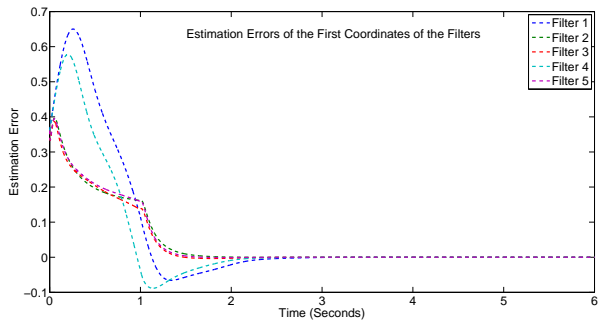}
\caption{The estimation errors of the $5$ filters in their first coordinates.}\label{fig1}
\end{center}
\end{figure}

To illustrate performance of the filters, they are simulated using MATLAB. The
simulation was implemented with the maximum step size of $0.001$ seconds and
simulation time of $10$ seconds. The system disturbances are modeled as pulse
signals of appropriate dimensions that last for $1$ seconds. The
initial state is drawn from a 
normally distributed vectors with mean of $0.1$ and standard deviation $0.2$.
  Figure~\ref{fig1} %and~\ref{fig2}
 shows the plot of the
 estimation error of the first coordinate of each filter, 
 $\hat{x}^1_i(t)-x^1(t)$. The plots confirm that
 despite some of local filters alone lead to an undetectable estimation
 problem, the 
 proposed distributed filter has successfully utilized imperfect communications
 between the nodes to provide estimates at each node with estimation
 errors converging to zero. %As a result the consensus  errors between the agents also converge to zero as time  evolves (Figure~\ref{fig2}).   
%   \begin{figure}
% \begin{center}
% \includegraphics[scale=.23]{Cons_Err.eps}
% \caption{{\bf The consensus errors of the estimation of each sensor and the rest of the sensor estimates, normalized with respect to the size of the plant state.}}
% \label{fig2}
% \end{center}
% \end{figure}
 
%  It is important to recognize that the conditions~\eqref{LMI} are sufficient to ensure the stability of the overall filter network. These conditions are conservative and as indicated by our simulations, even in cases that they are not satisfied, the estimation errors of the proposed filters do converge to zero by simply choosing $M^{-1}=0$. 
%=====================================================================
 \section{CONCLUSIONS}\label{conclusion}
%=====================================================================
 In this paper we proposed a distributed filtering algorithm by utilizing
 an $H_{\infty}$ minimum-energy filtering approach. The algorithm employs
 a decoupled computation of the individual filter coefficients. This is
 achieved by considering the estimation error of neighbouring agents as
 additional exogenous disturbances weighted according to the nodes'
 confidence in their neighbours' estimates. The proposed algorithm is shown
 to provide guaranteed internal 
 stability and desired disturbance attenuation of the network error
 dynamics. The paper has discussed tuning of the algorithm. We have also
 provided a simulation example that confirms convergence of the proposed
 algorithm in the case a Chua circuit with undetectable pairs $(A,C_i)$ at some of the nodes.  

%%%%%%%%%%%%%%%%%%%%%%%%%%%%%%%%%%%%%%%%%%%%%%%%%%%%%%%%%%%%%%%%%%%%%%%%%%%%%%%%

%%%%%%%%%%%%%%%%%%%%%%%%%%%%%%%%%%%%%%%%%%%%%%%%%%%%%%%%%%%%%%%%%%%%%%%%%%%%%%%%

%%%%%%%%%%%%%%%%%%%%%%%%%%%%%%%%%%%%%%%%%%%%%%%%%%%%%%%%%%%%%%%%%%%%%%%%%%%%%%%%

%%%%%%%%%%%%%%%%%%%%%%%%%%%%%%%%%%%%%%%%%%%%%%%%%%%%%%%%%%%%%%%%%%%%%%%%%%%%%%%%

\bibliographystyle{IEEEtran}
 \bibliography{ref}

\end{document}